\documentclass[copyright,creativecommons]{eptcs}
\providecommand{\event}{Gabriel Ciobanu (Ed.): Membrane Computing and\\ Biologically Inspired Process Calculi 2009} 
\providecommand{\volume}{15}   
\providecommand{\anno}{2009}   
\providecommand{\firstpage}{1} 
\providecommand{\eid}{1}       
\usepackage{latexsym}
\usepackage{graphicx}
\usepackage{textcomp}
\usepackage{amssymb}
\usepackage{marvosym}
\usepackage{amsmath}
\newcommand{\qqop}[1]{\mathrel{\makebox[2em]{$#1$}}}

\newcommand{\agr}{\quad\big|\quad}

\newcommand{\EE}{\mathcal{E}}
\newcommand{\DD}{\mathcal{D}}
\newcommand{\BB}{\mathcal{B}}

\newcommand{\RR}{\mathcal{R}}
\newcommand{\PP}{\mathcal{P}}
\newcommand{\Seq}{\mathcal{S}}
\newcommand{\XX}{\mathcal{X}}
\newcommand{\TT}{\mathcal{T}}

\newcommand{\ltrans}[1]{\xrightarrow{#1}}

\newcommand{\Loop}[1]{\ensuremath{\big(#1\big)^L}}

\newcommand{\into}{\ensuremath{\;\rfloor\;}}
\newcommand{\pipe}{\ensuremath{\;|\;}}

\newtheorem{definition}{Definition}

\title{Modelling Cell Cycle using Different Levels of Representation}

\author{Thomas Anung Basuki \institute{International Institute for Software Technology, United Nations University\\
    Macau SAR, China} \institute{Dipartimento di Informatica, Universit\`{a} di Pisa\\
      Largo B. Pontecorvo 3, 56127 Pisa, Italy} \email{anung@iist.unu.edu} \and
      Antonio Cerone \institute{International Institute for Software Technology, United Nations University\\
    Macau SAR, China} \email{antonio@iist.unu.edu} \and
      Rafael V. Carvalho \institute{International Institute for Software Technology, United Nations University\\
    Macau SAR, China} \email{rvcarvalho@iist.unu.edu}}

\def\titlerunning{Modelling Cell Cycle in Different Levels of Abstraction}
\def\authorrunning{T. A. Basuki \& A. Cerone \& R. V. Carvalho}

\begin{document}
\maketitle

\begin{abstract}
Understanding the behaviour of biological systems requires a
complex setting of in vitro and in vivo experiments, which
attracts high costs in terms of time and resources. The use
of mathematical models allows researchers to perform 
computerised simulations of biological systems, which are
called in silico experiments, to attain important insights
and predictions about the system behaviour with a 
considerably lower cost. Computer visualisation is an
important part of this approach, since it provides a 
realistic representation of the system behaviour. We define
a formal methodology to model biological systems using
different levels of representation: a purely formal
representation, which we call molecular level, models the
biochemical dynamics of the system; visualisation-oriented 
representations, which we call visual levels, provide views 
of the biological system at a higher level of organisation 
and are equipped with the necessary spatial information to
generate the appropriate visualisation. We choose Spatial
CLS, a formal language belonging to the class of Calculi 
of Looping Sequences, as the formalism for modelling all 
representation levels. We illustrate our approach using the 
budding yeast cell cycle as a case study.
\end{abstract}


\section{Introduction}

The high complexity of biological systems has encouraged during the last decades extensive inter-disciplinary research comprising biology and other fields of science such as computer science, mathematics, physics and chemistry. This interdisciplinary approach to biology resulted in a new field of study, called systems biology, which focuses on the systematic study of complex interactions in biological systems.

Computer scientists find many interesting similarities between systems biology and theory of concurrency. Degano and Priami~\cite{degpri03} claim that both systems biology and formal methods for concurrency can cross-fertilize each other. Being based on sound and deep mathematics, concurrency theories may offer solid ways to describe biological systems and safely reason upon them. On the other hand, systems biology studies many complex biological phenomena. Modelling and reasoning about these complex phenomena may require techniques that are more efficient and reliable than existing techniques. It is expected that the effort to understand biological mechanisms in terms of computer technology will possibly lead to new techniques that are more robust, efficient and reliable to model and analyse complex systems.

Many mathematical formalisms have been proposed to model and analyse systems biology. Some of them are based on formalisms generally used for modelling concurrent systems, such as Petri Nets~\cite{redliemav96,redmavlie93,harrob04}, the $\pi$-Calculus~\cite{chiacurdegmar05,priqua05,verbus07,verbus08}, and CCS (Calculus of Communicating Systems) ~\cite{dankri07}. Other formalisms are inspired by biological phenomena, such as compartmentalisation: 
Brane Calculi~\cite{cardelli05,danpra05}, P Systems~\cite{paunroz02,paun02}, Calculi of Looping Sequences~\cite{barcarmagmilpar08,SpatialCLS}, and Biocham ~\cite{Biocham06,Biocham05}. In particular, Calculi of Looping Sequences is a class of formalisms: 
Calculus of Looping Sequences (CLS) is the basic formalism~\cite{barcarmagmilpar08} and has two important extensions, 
Stochastic CLS~\cite{barcarmagmilpar08} in which reactions are associated with rates, and Spatial CLS~\cite{SpatialCLS}, which includes spatial information. 

These formalisms support the analysis of biological systems using tools based on numerical simulation, stochastic simulation and model-checking techniques. 
Bianco and Castellini developed PSim, a simulator for Metabolic P Systems, a variant of P Systems~\cite{PSim}. Scatena developed a simulator for Stochastic CLS~\cite{scatena}. 
Biocham is equipped with a tool for the simulation and model-checking of biological systems. The probabilistic model-checker PRISM has also been used to analyse some biological properties~\cite{heaetal03,KNP08,heaetal08}. 
All these tools provide textual output as well as a plot of the simulation.

Texts and plots provide very detailed information on specific aspects of the analysed biological system. However, they are often inadequate when the aim is to acquire global knowledge about the high-level organisation and dynamics of the biological system. For example, in most experiments the analyst can only vary molecular concentrations in the environment and within cells, whereas the aim of an experiment or simulation may be to observe the resultant behaviour of cells or even the whole organ or organism. Such high-level behaviours can be better described through two or three dimensional visualisation/animation rather than using texts and plots.

One approach for modelling and visualising biological systems is based on the use of L-systems. L Systems use rewriting systems for modelling biological processes~\cite{Gia04}. However, they are used mainly for visualising the development of plants~\cite{HamPru96,Pow99,GAU00}. Hammel and Prusinkiewicz model the behaviour of Anabaena at cellular level~\cite{HamPru96} by considering interactions between cells with external factors in the environment.

Michel, Spicher and Giavitto use rule-based programming language MGS to model and simulate the $\lambda$ phage genetic switch~\cite{Mic09}. They present a multilevel model of the system; a molecular level defined using built-in Gillespie's algorithm and a population of cells level defined using GBF (Group Based Field) and Delaunay topological collections.

David Harel and his group developed an approach in modelling at different levels of representation~\cite{harelvisual}.
They use object oriented approach and define the cell as the basic building block of their approach. Their approach uses scenario to define system behaviour and uses animation on a 2-dimensional grid~\cite{gemcell}. Scenarios define cell behaviour related with interactions between molecules in the environment and their receptors on cell membranes. Another interesting application of their approach is the modelling of pancreatic organogenesis~\cite{harelpancreas}. In this application they show how molecular interactions affect cell growth and, in the end, affect the growth of mammalian pancreas. A three dimensional visualisation is used to visualise the pancreatic organogenesis process.

Another tool used to visualise biological systems according to a model of the system at the molecular level is Virtual Cell~\cite{virtualcell}. This tool is based on a deterministic numerical simulation of the model, which is defined using differential equations.



Technology has helped biologists to observe biological systems at microscopic level, down to the molecular level. More knowledge has been gained about cell structure and behaviour. Although it is possible to track the causes of certain phenomena at cellular level down to specific biochemical reaction occurring at molecular level, we are still far from being able to entirely explain cell behaviour in terms of the biochemical reactions occurring within the cell and its environment. Moreover, based on natural observation and experiments, it is also possible to give an accurate description of all phenomena observable at cellular level.

We define an approach to model biological systems at different levels of representation. We consider the molecular level, in which organic molecules interact through biochemical reactions, as the lowest level of our hierarchy. At this level, the representation is merely a mathematical formalisation of biochemical reactions, with no visualisation. The representation of higher levels of organisation and dynamics such as a cell, a tissue, an organ or an organism, is inspired by observations of the system behaviour under normal conditions. At these levels the mathematical formalisation mimics organisation and dynamics observed in nature and replicated in controlled experiments, with the aim to give a visualisation, possibly in three dimensions, of the modelled phenomena. We refer to these higher levels as visual levels, in contrast to the molecular level, for which we do not provide any visualisation. Each visual level has to be linked to the molecular level by a formal description of the way biochemical reactions cause transitions of visual states. However such causal relations are not always fully understood in terms of biological theories. Therefore, transitions of visual states may sometimes be governed by rates observed in nature rather than by the underlying biochemistry modelled at molecular level or may be directly associated with the introduction or removal of biochemical signals, whose accumulation or degradation is not sufficiently explained in terms of biological theories.

This approach allows us to introduce changes to cells and their environment at the molecular level and observe the impact of such changes at the visual level. Changes at the molecular level may range from varying the concentration of a specific molecule or introducing a new kind of molecule with specific properties to the introduction of a plasmid within the cell or a virus in its environment. 

The work of David Harel and his group~\cite{harelvisual,gemcell} and the work of Hammel and Prusinkiewicz~\cite{HamPru96} have similarity with our approach, in the sense that they also represent both molecular and cellular level. However their approaches consider molecules only as external factors that affect cells behaviour. The cellular level is modelled only in the environment and on the cell membrane, but not inside the cell. 
Moreover in David Harel's approach, all rules are deterministic and randomness can only be introduced by defining random initial states of the system. In contrast, in our approach, we model and simulate stochastic behaviour by introducing reaction rates and controlling the occurrences of reactions using Gillespie's algorithm.

Slepchenko, Schaff, Macara and Loew take an approach closer to ours by developing a tool to model biological systems at molecular level and visualise them at cellular level~\cite{virtualcell}. Their approach, however, is based on differential equations, which are deterministically simulated using numerical simulation methods.

In this paper we use Spatial CLS to model both the molecular  and visual level. We consider the cellular level as visual level and we use the cell cycle of budding yeast as a case study. However, spatial information is only introduced at the visual level, while the molecular level is modelled using a subset of Spatial CLS with no spatial information, which is equivalent to Stochastic CLS.

The simulation is performed using a variant of Gillespie's algorithm which extends the variant introduced by Basuki, Cerone and Milazzo~\cite{SCLSMaude} with the addition of a mechanism for choosing individual cells, which is governed by exponential distribution of reaction time.


\section{Using Spatial CLS for Visualisation}\label{CLS}
Calculi of Looping Sequences is a class of formalisms developed at the University of Pisa for the purpose of modelling biological systems~\cite{milazzo,barcarmagmilpar08}. In this section we focus on one variant of the calculi called Spatial CLS~\cite{SpatialCLS}, which supports the modelling of details about position and movement of the components in the system.

A Spatial CLS model contains a term and a set of rewrite rules. The term describes the initial state of the system, and the rewrite rules describe the events which may cause the system to evolve. We start by defining the syntax of terms. We assume a possibly infinite alphabet $\EE$ of symbols ranged over by $a,b,c,\ldots$ and a set $\mathcal{M}$ of names of movement functions. Each symbol represents an atomic component of the system.
\begin{definition} \emph{Terms} $T$, \emph{Branes} $B$ and \emph{Sequences} $S$ are
given by the following grammar:
\begin{eqnarray*}
T & \qqop{::=} & \lambda \agr (S)_d \agr \Loop{B_d} \into T \agr T \pipe T \\
B & \qqop{::=} & \lambda \agr (S)_d \agr B \pipe B \\
S & \qqop{::=} & \epsilon \agr a  \agr S \cdot S
\end{eqnarray*}
\normalsize
where $a$ is a generic element of $\EE$, $\epsilon$ represents the empty sequence, and $d \in \DD = ((\mathbb{R}^n\times\mathcal{M})\cup\{.\})\times\mathbb{R}^+$. We denote with $\TT$, $\BB$ and $\Seq$ the infinite set of terms, branes and sequences.
\end{definition}

There are four operators in the formalism. A sequencing operator $\_\cdot\_$ is used to compose some components of the system in a structure that has the properties of a sequence. For instance, sequencing can be used to model DNA/RNA strands or proteins. The looping operator $\Loop{\_}$ and containment operator $\_\into\_$ are always applied together, hence they can be considered as a single binary operator $\Loop{\_} \into \_$ to be applied to one brane and one term. Looping and containment allow the modelling of membranes and their contents. Finally the parallel composition operator $\_\pipe\_$ is used to compose juxtaposition of entities in the system. Brackets can be used to indicate the order of application of the operators, and we assume $\Loop{\_} \into \_$ to have precedence over $\_\pipe\_$.

In Spatial CLS there are two kinds of terms, positional and non-positional terms. Positional terms have spatial information, while non-positional terms do not. A term with a `.` in its spatial information represents a non-positional term.

Every positional term is assumed to occupy a space, modelled as a sphere. The spatial information of a term contains three parts: the position of the centre of the sphere, its radius and a movement function. In this way every object is assumed to have an autonomous movement. Variable $d$ in the above syntax models such spatial information. In this paper we only consider terms without autonomous movement. Therefore, we can omit movement functions from our spatial information.

We now define patterns, which are terms enriched with variables. We assume a set of position variables $PV$, a set of symbol variables $\XX$ and a set of sequence variables $SV$. We denote by \textbf{T} the set of all instantiation functions $\tau : PV \rightarrow \DD$ for the position variables.

\begin{definition} \emph{Left Brane Patterns} $BP_L$, \emph{Sequence Patterns} $SP$ and \emph{Right Brane Patterns} $BP_R$ are given by the following grammar:

$
\begin{array}{rcl}
BP_L & \qqop{::=} & (SP)_u \agr BP_L \pipe BP_L \\
BP_R & \qqop{::=} & (SP)_g \agr BP_R \pipe BP_R \\
SP & \qqop{::=} & \epsilon \agr a \agr SP . SP \agr \tilde{x} \agr x
\end{array}
$

where $u\ \in\ PV$, $x\ \in\ \XX$, $\tilde{x}\ \in\ SV$ and $g \in \mathbf{T}$
\end{definition}

\begin{definition} \emph{Left Patterns} $P_L$ and \emph{Right Patterns} $P_R$ are given by the following grammar:
\begin{eqnarray*}
P_L & \qqop{::=} & (SP)_u \agr \Loop{BP_{LX}}_u \into P_{LX} \agr P_L \pipe P_L\\
BP_{LX} & \qqop{::=} & BP_L \agr BP_L\pipe\bar{X} \agr \bar{X}\\
P_{LX} & \qqop{::=} & P_L \agr P_L\pipe X\\
P_R & \qqop{::=} & \epsilon \agr (SP)_g \agr \Loop{BP_{RX}}_g \into P_R \agr P_R\pipe P_R\agr X\agr \bar{X} \\
BP_{RX} & \qqop{::=} & BP_R \agr BP_R\pipe\bar{X} \agr \bar{X}
\end{eqnarray*}
\normalsize
where $u\ \in\ PV$, $x\ \in\ \XX$, $\tilde{x}\ \in\ SV$ and $g \in \mathbf{T}$.
\end{definition}

We denote the sets of all left patterns by $\PP_L$, the set
of all right patterns by $\PP_R$. We denote by Var($P$) the set of all variables appearing in a pattern
$P$, including position variables from $PV$.

\begin{definition} A \emph{rewrite rule} is a 4-tuple ($f_c, P_L, P_R, k$), usually written as
\[[f_c] P_L \stackrel{k}{\mapsto} P_R\]
where $f_c : \mathbf{T} \rightarrow \{tt,f\! f \}$, $k \in R^+$, Var($P_R$) $\subseteq$ Var($P_L$), and each function g appearing in $P_R$ refers only to position variables in Var($P_L$).
\end{definition}

Rewrite rules are used to define reactions that may occur in a system. The left and right patterns of a rule is matched against the term that represents the current state of the system, to check its applicability. It is possible to define a rewrite rule that has a precondition defined by $f_c$. A precondition is checked against any term matches in the left pattern of a rule. The rate constant $k$ models the propensity of a reaction. The value 1/$k$ represents the expected duration of a reaction involving reactant combinations.

Barbuti, Maggiolo-Schettini, Milazzo and Pardini~\cite{SpatialCLS} define a semantics for Spatial CLS, as a Probabilistic Transition System. In this semantics, they consider each evolution step of a biological system as composed of two phases: (1) application of at most one reaction; (2) updating of positions according to the movement functions. In this paper we omit movement functions, simplifying the evolution step into one phase only.

The combination of looping and containment operators and parallel composition operators in Spatial CLS defines a notion of layers in the terms. The parallel composition operators model the objects as multisets, while the looping and containment operators create boundaries between these multisets. Objects within the same boundary are considered to be in the same layer. We need to define some functions in order to calculate the rate of a rule application. In the following definitions, we denote the multiset of top-level elements appearing in a pattern P by \={P}, and assume the function $\mathbf{n} : \TT \times \TT \rightarrow \mathbb{N}$ such that $\mathbf{n}(T_1,T_2)$ gives the number of times $T_1$ appears at top-layer in $T_2$. We define $\tau$ as instantiation function for position variables and $\sigma$ as instantiation function for other variables.

$
\begin{array}{rlr}
comb(P_{L1}\pipe P_{L2},\tau,\sigma) = & comb(P_{L1},\tau,\sigma) . comb(P_{L2},\tau,\sigma) & \\
comb(\Loop{BP_{LX}}_u \into P_{LX},\tau,\sigma) = & comb'(BP_{LX},\tau,\sigma) . comb'(P_{LX},\tau,\sigma) & \\
comb(SP_u,\tau,\sigma) = & 1 & \\
comb(P_L\pipe U,\tau,\sigma) =
& \prod_{T\in \bar{P_L\tau\sigma}} \left( \begin{array}{l}
                                                        \mathbf{n}((PL|U)\tau\sigma, T )\\
                                                        \mathbf{n}(P_L\tau\sigma, T ) \end{array} \right) . comb(P_L,\tau,\sigma), &
                                                        U \in BV \cup TV \\
comb'(P_L,\tau,\sigma) = & comb(P_L,\tau,\sigma) & \\
binom(T_1,T_2,T_3) = & \prod_{T\in \bar{T_1}} \prod_{i=1}^{\mathbf{n}(T_3,T)}
                                                \frac{\mathbf{n}(T_2,T)+i}{\mathbf{n}(T_2,T)-\mathbf{n}(T_1,T)+i}& \\
\end{array}
$
\normalsize

Given a finite set of rewrite rules $\RR$, let $\RR_B \subseteq R$ be the set of all brane rules and let $\stackrel{R,T,c}{\rightarrow}$
 with $R \in \RR$, $T \in \TT$ and $c \in \mathbb{N}$, be the least labeled transition
relation on terms satisfying inference rules in Figure~\ref{Semantics}.

\begin{figure}[!htb]
    \begin{center}
        $
\begin{array}{cccccc}
\multicolumn{2}{c}{(R: [f_c] P_L\stackrel{k}{\mapsto} P_R) \in \RR} & & f_c (\tau) = tt & \tau \in \mathbf{T} & \sigma \in \sum \\ \hline
& \multicolumn{4}{c}{P_L\tau\sigma \ltrans{R,P_L\tau\sigma,comb(P_L,\tau,\sigma)} P_R\tau\sigma} & \\ \\
B \ltrans{R,T,c} B' & R \in \RR_B & & & \multicolumn{2}{l}{T_1 \ltrans{R,T,c} T'_1} \\ \cline{1-2} \cline{4-6}
\multicolumn{2}{c}{\Loop{B}_d \into{T_1} \ltrans{R,\Loop{B}_d \into{T_1},c} \Loop{B'}_d \into{T_1}} &
& \multicolumn{3}{c}{\Loop{B}_d \into{T_1} \ltrans{R,\Loop{B}_d \into{T_1},c} \Loop{B'}_d \into{T_1}} \\ \\
& \multicolumn{3}{c}{T_1 \ltrans{R,T,c} T'_1} & & \\ \cline{2-4}
& \multicolumn{3}{c}{T_1\pipe T_2 \ltrans{R,T,c.binom(T,T_1,T_2)} T'_1\pipe T_2} & &
\end{array}
$
        \caption{Inference rules used for calculating rates of rewrite rules}
       \label{Semantics}
    \end{center}
\end{figure}

The following definition gives all the reactions enabled in a state, by also taking
into account the subsequent rearrangement:
\[ Appl(R, T) = {(Tr, c, T', T'') | T \ltrans{R,T,c}  T' \land T'' = Arrange(T') \neq \perp} \]

The number $m_T^{(R)}$ of different reactant combinations enabled in state T, for
a reaction $R$, and the total number $m_T$ of reactions considering a set of rules $\RR$,
are defined as:
\[ m_T^{(R)} = \sum_{(Tr,c,T',T'')\in Appl(R,T)} c\] and \[m_T = \sum_{R\in \RR} m_T^{(R)}\]

Let $T$ describe the state of the system at a certain step, and $k_R$ denote the rate
associated with a rewrite rule $R$. At each step of the evolution of the system, in
order to assume that at most one reaction can occur, we have to choose a time
interval $\delta t$ such that $(\sum_R\in \RR k_R m_T^{(R)})\delta t  \leq 1$. Given a set of rewrite rules $\RR$, we
choose an arbitrary value $N$ such that for each rule $R \in \RR$ it holds $0 < kR/N \leq 1$.
Then we compute the time interval for a step as $\delta t = 1/Nm_T$, thus satisfying the above
condition. The value of $N$ also determines the maximum permitted length of each
step as $1/N$ time units.

The probability that no reaction happens in the time interval $\delta t$ is:
\[\bar{p}_T = 1 - \sum_{R\in \RR} (\sum_{(Tr,c,T',T'')\in Appl(R,T)} \frac{k_R}{Nm_T}c)\]
and the probability $P(T_1 \rightarrow T_2, t)$ of reaching state $T_2$ from $T_1$ within a time interval
$\delta t $ after $t$ is such that:
\[P(T_1 \rightarrow T_2, t) = \sum_{R\in \RR} (\sum_{(Tr,c,T',T_2)\in Appl(R,T_1)} \frac{k_R}{Nm_{T_1}}c)+
\left\{ \begin{array}{ll} \bar{p}_{T_1} & \mbox{if $T_1 = T_2$}\\
                          0 & \mbox{otherwise} \end{array}\right.\]

The semantics of Spatial CLS is defined as Probabilistic Transition System as follows.
\begin{definition} Given a finite set of rewrite rules $\RR$, the semantics of Spatial CLS is the least relation satisfying the following inference rules:
\\
$
\begin{array}{p{0.5in}ccc}
& (T_r, c, T', T_2) \in Appl(R, T_1) & & R \in \RR  \\
& p = P(T_1 \rightarrow T_2 , t) & & \delta t = \frac{1}{Nm_{T_1}} \\ \cline{2-4}
& & \langle T_1, t\rangle \stackrel{p}{\rightarrow}
\langle T_2, t + \delta t\rangle & \\ \\
& p = P(T \rightarrow T' , t) & & \delta t = \frac{1}{N\ max(1,m_{T_1})} \\ \cline{2-4}
& & \langle T, t\rangle \stackrel{p}{\rightarrow}
\langle T', t + \delta t\rangle &
\end{array}
$
\end{definition}

Application of a reaction may still result in non well-formed terms (e.g. a term that contains collision of two objects). To solve this problem, we define the algorithm described in Section \ref{Cell Cycle} to rearrange objects in the system. In this section we describe some assumptions about space and spatial information used in our approach.

First, we limit the direction where any object in the system can move. We assume that the space occupied by each object in the system is in $\mathbb{R}^n$ and each object can only move along one axis at one time. For every axis, there are two possible directions. So in total there are $2n$ possible directions. Therefore in $\mathbb{R}^3$ any object can move along 6 possible directions.

Positional terms in Spatial CLS have size, which is defined as the radius of the sphere that encloses the term. It is possible to define reactions that modify this size. In our algorithm, we assume that the maximum sizes of all objects in the same layers are known. 
We assume the existence of an $n$-dimensional grid, which divides the space occupied by a biological system into cubes (we are working in n=3 dimensions) with fixed size defined in such a way that every object in the system fits inside it. 

Initially objects of a biological system are positioned inside cubes. Reactions defined for a biological system can create new objects, move existing objects to different positions or remove objects from the system. New objects are also positioned inside cubes. 
Therefore, the only time we need rearrangement is when two objects are inside the same cube. In this case, we will need to run the rearrangement algorithm explained in Section \ref{Cell Cycle}.





\section{Levels of Representation in Spatial CLS}

In this section we define an approach to model biological systems at different levels of representation using Spatial CLS. We distinguish between a molecular level, in which rewrite rules are used to model biochemical reactions among molecules, and one or more visual levels, in which rewrite rules define the dynamics of a higher level of organisation of the biological system under analysis. 
These rewrite rules refer to a single level of representation. Therefore, we call them \textit{horizontal rules}.

In this paper we consider only the cellular level as a visual level. At visual level the state of the system, which is called visual state, is defined using the spatial information in positional terms of  the system. A visual state describes three kinds of information:
\begin{enumerate}
    \item spatial information;
    \item a stage of the system evolution, which we call visual stage;
    \item information on whether that stage has been visualised.
\end{enumerate}
Example of stages of the system at cellular level are a small cell at the beginning of the growing phase or a cell with two nuclei during mitosis. Within a specific stage, state transitions are triggered by rewrite rules that modify spatial information and tag the current stage of the system evolution as ``visualised''.

In this way, we  can attain visualisation using spatial information. Moreover, visual states represent both the biological stage of the system and the status of the visualisation, 
while rewrite rules control the flow of the visualisation. Visualisation can then be used to simulate the behaviour of the system and to perform comparison with and prediction of in vivo and in vitro experiments.

At molecular level we can see biological systems as composed of molecules which are not associated with spatial information in our model. The state of the system is represented by the combination of molecular populations. State transitions are defined as rewrite rules representing biochemical reactions modifying molecular populations. 
Therefore the molecular level is modelled using a subset of Spatial CLS equivalent to Stochastic CLS.

To model a biological system, we can start from the visual level. The information about the system behaviour at this level is purely descriptive. It is based on observation of visible events independently of the biochemical processes that cause them. Such visual events are modelled through transitions of visual states, whose spatial information is defined in such a way to mimic visible events observed during in vivo and in vitro experiments. 
Molecules are confined within membranes using the looping and containment operator. However, such a confinement is logical rather than spatial. Chemical reactions are modelled by rules with no visual effect and then linked to the visual level by \textit{vertical rules}, which 
control the transition of stage in the system evolution, by evaluating conditions at molecular level and checking that the current stage has been visualised.

\section{Case Study: Cell Cycle}\label{Cell Cycle}

In this section we illustrate our approach using the cell cycle  of budding yeast as a case study.

\subsection{Visual Level}\label{visual}
Cell cycle consists of four phases: $G_1$ - $S$ - $G_2$ - $M$. 
However, at cellular level, only phase $S$ fully characterises a visual stage, that is the stage where chromosomes inside the nucleus are replicated. Phase $G_1$ incorporates different steps of cell growth, phase $G_2$ does not have any visual counterpart and phase $M$ includes one visual stage corresponding to nucleus division.

In our model we consider only two steps in cell growth: the cell size before and after the growth. Therefore we define 4 visual stages:
\begin{enumerate}
    \item small cell before growth (beginning of phase $G_1$);
    \item big cell after growth (end of phase $G_1$);
    \item chromosomes inside the nucleus (end of phase $S$);
    \item cell with two nuclei (phase $M$ before cytokinesis).
\end{enumerate}

Based on the above explanation, we define three variables identifying visual stages:\begin{itemize}
    \item cell radius;
     \item number of nuclei in a cell;
     \item number of chromosomes in a cell nucleus.
\end{itemize}
Table \ref{tab:visualstages} shows the values of these variables in each visual stage.

\begin{table*}
    \centering
        \begin{tabular}{||c|c|c|c|c||} \hline
            Stage & cell radius & \# of nuclei & \# of chromosomes & avg. time (min) \\ \hline
            1 & $3r/4$ & single & single & 40 \\ \hline
            2 & $r$ & single & single & 30 \\ \hline
            3 & $r$ & single & double & 25 \\ \hline
            4 & $r$ & double & double & 5 \\ \hline
        \end{tabular}
        \caption{The four visual stages of cell-cycle}\label{tab:visualstages}
\end{table*}

State transitions are defined using Spatial CLS rewrite rules.
We adopt an example similar to the one presented by Barbuti, Maggiolo-Schettini, Milazzo and Pardini~\cite{SpatialCLS}. 
We define four rewrite rules to model the cell cycle. Barbuti, Maggiolo-Schettini, Milazzo and Pardini consider the 24 hour mammalian cell cycle~\cite{SpatialCLS}. In this paper, we model budding yeast cell cycle whose duration is only about 100 minutes~\cite{Chenetal04}. The initial state of the system is defined by the following term:
\[(b)^L_{.,R}\ \rfloor\ (m)^L_{[0,0,f],\frac{3r}{4}}\ \rfloor\ ((n)^L\ \rfloor\ (cr . gN2 . gB5\,|\, cr . gB2 . gC20) | stage_1)  \]
The above term represents a sphere with radius $R$, which contains a cell positioned in its centre. The cell contains one nucleus, with 2 chromosomes inside. Each chromosome contains 2 genes, whose function will be explained in Section \ref{molecular}. The cell is initially in stage 1 (phase $G_1$). At this level the cell cycle is defined by the following rules:

\small
\begin{flushleft}
\begin{description}
\item $R_1 : \Loop{m}_{[p,f],\frac{3r}{4}} \into{(X\ \pipe\ stage_1)} \stackrel{0.025}{\longmapsto} \Loop{m}_{[p,f],r} \into{(X\pipe stage_1\pipe visualised_1)}$

\item $R_2 : \Loop{m}_{[p,f],r} \into{((n)^L_u\ \rfloor\ (cr . \tilde{x}\, |\, cr . \tilde{y})\ \pipe\ stage_2)}  \stackrel{0.033}{\longmapsto} (m)^L_{[p,f],r}\ \rfloor\ ((n)^L_u\ \rfloor\ (2cr . \tilde{x}\ |\ 2cr . \tilde{y})\ \pipe\ stage_2\ \pipe visualised_2)$

\item $R_3 : \Loop{n}_{[(0,0,0),f],\frac{2r}{5}}\ \rfloor\ (2cr . \tilde{x}\ |\ 2cr . \tilde{y})\ |\ stage_3 \stackrel{0.04}{\longmapsto} (n)^L_{[(-\frac{r}{2},0,0),f],\frac{2r}{5}}\ \rfloor (cr . \tilde{x}\ |\ cr . \tilde{y})\ |\ \Loop{n}_{[(\frac{r}{2},0,0),f],\frac{2r}{5}}\ \rfloor\ (cr . \tilde{x}\ |\ cr . \tilde{y})\ |\ stage_3\ |\ visualised_3$

\item $R_4 : \Loop{m}_{[p,f],r}\ \rfloor\ (\Loop{n}_u\ \rfloor\ X\ |\ \Loop{n}_v\ \rfloor\ Y\pipe\ stage_4) \stackrel{0.2}{\longmapsto} \Loop{m}_{[p,f],\frac{3r}{4}}\ \rfloor\ (\Loop{n}_u\ \rfloor\ X\ |\ stage_4\ \pipe visualised_4)\ | \Loop{m}_{[get\! pos,f],\frac{3r}{4}}\ \rfloor\ (\Loop{n}_u\ \rfloor\ Y\ \pipe\ stage_4\ \pipe visualised_4)$
\end{description}
\end{flushleft}
\normalsize

In the above rules we only model objects without autonomous movement. Function $f$ on their spatial information represents a function that maps from position $p$ to the same position $p$.
Every cell is assumed to double its volume during cell cycle. This is shown by rule $R_1$, which represents the growing process in phase $G_1$. By changing the cell radius from $\frac{3r}{4}$ to $r$, the volume is nearly doubled. Rule $R_2$ represents the chromosomes replication inside nucleus. It is represented by modifying symbol $cr$ that precedes each chromosome to $2cr$. Rule $R_3$ represents the nucleus division, where the only nucleus inside a cell is duplicated into two identical nuclei. To avoid collision between nuclei, pairs of nuclei are moved toward opposite directions. Finally rule $R_4$ represents the cytokinesis, which divides the cell and all its contents into two daughter cells with the same size and content. This is represented by (1) removing the mother cell whose radius is $r$ and having two nuclei, (2) putting a daughter cell with radius $\frac{3r}{4}$ at mother cell's position, 
(3) putting a daughter cell with radius $\frac{3r}{4}$ at a new position determined by $get\! pos()$. 
Symbols $stage_1, stage_2, stage_3, stage_4$ are used to model the current visual stage of a cell. Symbols $visualised_1, visualised_2, visualised_3, visualised_4$ decompose each visual stage into two visual states, which model whether the current stage has been visualised or not. Therefore, at visual level, rewrite rule $R_i$ models the transition between the two visual states that correspond to visual stage $i$, that is from the non-visualised to the visualised state of stage $i$. 

Although in general rewrite rules at visual level modify spatial information, this is not the case for rule $R_2$ because chromosomes are logically positioned within the nucleus, but do not have any quantitative spatial information associated with them. In fact rule $R_2$ just double the number of chromosomes; choices about where to visualise the duplicated chromosomes are purely aesthetic and are left to the implementation of the visualisation.

The constant associated with each rule defines the rate with which that rule is applied. The rate for rule $R_i$ is calculated as the inverse of the average duration for visual stage $i$. This ensures that the time needed for each stage to be visualised mimics the actual time observed in nature. The last column of Table~\ref{tab:visualstages} shows the average durations of the four stages.

\subsection{Molecular Level and Vertical Rules}\label{molecular}

To model cell cycle at molecular level, we adopt the model of budding yeast cell cycle introduced by Chen et al.~\cite{Chenetal04}. It is a very detailed model based on differential equations. There is also a variant of this model for the eukaryotic cell cycle~\cite{CsiNag06}. Li et al. developed a simpler model of budding yeast cell cycle  using boolean network~\cite{Li04}.

Cell cycle is controlled by complexes of cyclin and cyclin-dependent kinases. There are 4 kinds of cyclins involved in cell cycle:\begin{itemize}
    \item \emph{Cln3}, which starts phase $G_1$;
    \item \emph{Cln2}, which induce the $G_1$/$S$ transition;
    \item \emph{Clb5}, which controls the phase $S$;
    \item \emph{Clb2}, which controls the phase $M$.
   \end{itemize}
These four cyclins form complexes with \emph{Cdc28}.

We define rules that control the state of the system at molecular level. To define a link between the visual state (which is controlled by the horizontal rules at visual level) and the biochemical state at molecular level, we define 4 vertical rules that cause a transition to next visual stage when a specific condition at molecular level is verified. Since these rules do not correspond to any time-consuming biochemical process but operate at a meta-level by providing a link between distinct representation levels, we define them as instantaneous by using $\infty$ as the value for their rates.

Let $cond_i$ be the condition at molecular level that triggers a transition from visual stage $i$ to next visual stage. We define vertical rule
\[ T_i : visualised_i | cond_i | stage_i \stackrel{\infty}{\longmapsto} cond_i | stage_{next(i)}\]
where \[ next(i) = \left\{
                                                    \begin{array}{ll}
                                                    1 & \mbox{if $i = 4$} \\
                                                    i+1 & \mbox{if $0 < i < 4$}
                                                    \end{array}
                                     \right. \]
Symbol $visualised_i$ is introduced by the application of rule $R_i$, which marks the completion of the visualisation of stage $i$. The completion of the visualisation of the current stage is obviously a precondition for the transition to next visual stage. Therefore, symbol $visualised_i$ appears as a precondition in vertical rule $T_i$, which defines the transition to next visual stage $next(i)$. The removal of $visualised_i$ by rule $T_i$ enables rule $R_{next(i)}$ to be applied.

Vertical rules also allow the introduction and removal of biochemical signals whose accumulation or degradation is not sufficiently understood in order to be dealt with at molecular level. The form of vertical rules that deal with introduction and removal of biochemical signals is as follows:
\[ \bullet\ \mbox{introduction,}\ T_i : visualised_i | cond_i | stage_i \stackrel{\infty}{\longmapsto} cond_i | stage_{next(i)} | signal\]
\[ \bullet\ \mbox{removal,}\ T_i : signal | visualised_i | cond_i | stage_i \stackrel{\infty}{\longmapsto} cond_i | stage_{next(i)}\]

We use the following convention for naming objects. Molecule names start with capital letters, except in some cases where the molecules could be in two different statuses. For example, we prefix the molecule names with $i$ to indicate that this molecule is in inactive status. We also prefix the molecule names with $p$ when the molecule is in phosphorylated status. Names starting with $g$ are used for genes. We use '-' to concatenate two names of molecules, indicating a complex formed by binding two molecules.

Cell cycle starts when a cell grows in phase $G_1$. This is triggered by a growth factor, which is present in the environment, and binds with its receptor in the cell membrane. The resultant complex then triggers the production of cyclin \emph{Cln3}. Cyclin \emph{Cln3} (after binding with its kinase partner \emph{Cdc28}) activates \emph{SBF} and \emph{MBF}, the transcription factors for cyclins \emph{Cln2} and \emph{Clb5}. Genes $gN2$ and $gB5$ control expression of cyclin \textit{Cln2} and \textit{Clb5}. Cyclin \emph{Cln2} controls the transition between phase $G_1$ and $S$. Cyclin \emph{Clb5} controls phase \emph{S}. At this point some molecules that are not needed in this phase, but will be needed in later phases, are temporarily deactivated. In phase $G_1$, \emph{Clb5} is deactivated by \emph{Sic1}. Later \emph{Cln2} forms a complex with \emph{Cdc28} and phosphorylates \emph{Sic1}, releasing \emph{Clb5}. \emph{Clb2}, which is a cyclin needed in mitosis, is bound with \emph{Sic1} in phase $G_1$. \emph{Cdc14}, which is needed in mitosis exit, is bound with \emph{Net1} in phase $G_1$. \emph{Cdc14}, which is abundant in this phase also activates phosphorylated \emph{Sic1}, the \emph{Clb5} inhibitor.

Based on their duration, we classify reactions into four categories: very fast, fast, slow and very slow. We define four numerical values to characterise reaction rates for the categories above: 20, 5, 1, 0.25. The higher the rate, the faster the reaction. Obviously, reaction times are many magnitudes smaller than durations of visual stages defined in Table~\ref{tab:visualstages}. Therefore, we introduce a speeding factor $s$, which defines the ratio between these magnitudes for the specific visual representation we are modelling. The actual rate of a reaction is then given by the product between the speeding factor and the numerical value corresponding to the category of that reaction.

To simplify the model, we don't define rules for complexation of cyclins with its kinase partners (\emph{Cdc28}). Phase $G_1$ is therefore modelled as follows:

\small
$
\begin{array}{ll}
    S_1 : & GF\, |\, (GFR\, |\, Y)^L\rfloor X \stackrel{20\cdot s}{\longmapsto} (iGFR\, |\, Y)^L\rfloor X\ |\ Cln3 \\
    S_2 : & Cln3\,|\, iSBF\, |\, iMBF \stackrel{1\cdot s}{\longmapsto} Cln3\, |\, SBF\, |\, MBF \\
    S_3 : & SBF\, |\, (n)^L\rfloor (\tilde{y}.gN2.\tilde{x}\, |\, Y) \stackrel{0.25\cdot s}{\longmapsto} SBF\, |\, (n)^L\rfloor (\tilde{y}.gN2.\tilde{x}\, |\, Y)\ |\ Cln2  \\
    S_4 : & MBF\, |\, (n)^L\rfloor (\tilde{y}.gB5.\tilde{x}\, |\, Y) \stackrel{0.25\cdot s}{\longmapsto} MBF\, |\, (n)^L\rfloor (\tilde{y}.gB5.\tilde{x}\, |\, Y)\ |\ Clb5 \\
    S_5 : & Sic1\ |\ Clb5 \stackrel{5\cdot s}{\longmapsto} Sic1-Clb5 \\
    S_6 : & Net1\ |\ Cdc14 \stackrel{5\cdot s}{\longmapsto} Net1-Cdc14 \\
    S_7 : & pSic1\ |\ Cdc14 \stackrel{20\cdot s}{\longmapsto} Sic1\, |\, Cdc14 \\
    S_8 : & Sic1\, |\, Clb2 \stackrel{5\cdot s}{\longmapsto} Sic1-Clb2 \\
    S_9 : & Cln2\ |\ Sic1-Clb5 \stackrel{5\cdot s}{\longmapsto} pSic1 \ |\ Clb5\ |\ Cln2 \\
\end{array}
$

\normalsize

The accumulation of \emph{Cln2} triggers the transition from phase $G_1$ to $S$~\cite{Chenetal04}. We introduce the following vertical rule:
\small
\[T_1 :  visualised_1 | Cln2^{mc(Cln2,2)} | stage_1 \stackrel{\infty}{\longmapsto} Cln2^n | stage_2 \]
\normalsize
to instantaneously perform the transition from stage 1 to stage 2, after rule $R_1$ has visualised the cell growth.
The function $mc(r,i)$ represents the minimum concentration of $r$ that is needed to trigger the transition to stage $i$.

Cyclin \emph{Clb5} forms a complex with its kinase partner \emph{Cdc28}, and triggers the duplication of chromosomes. Meanwhile \emph{Cln2} is not needed anymore, so it is degraded by \emph{SCF}. Protein \emph{SCF} also degrades phosphorylated \emph{Sic1}, enabling \emph{Clb5} to optimally work. The cyclin-dependent kinases (\emph{Cln2-Cdc28} and \emph{Clb5-Cdc28}) activate \emph{Mcm1}, which is the transcription factor for \emph{Clb2}. Expression of Clb2 is controlled by gene $gB2$. Cyclin \emph{Clb2} controls phase $M$. Active \emph{Cdh1}, whose function is to degrade \emph{Clb2}, is deactivated by cyclin-dependent kinases. In this way the degradation of \emph{Clb2} is postponed until the end of phase $M$. Finally \emph{Clb2} is transcribed and then binds with \emph{Cdc28}.

Phase $S$ is therefore modelled as follows:

\small
$
\begin{array}{ll}
    S_{10} : & pSic1\ |\ SCF \stackrel{1\cdot s}{\longmapsto} SCF \\
    S_{11} : & Cln2\ |\ SCF \stackrel{1\cdot s}{\longmapsto} SCF\ \\
    S_{12} : & Cln2\ |\ Cdh1 \stackrel{20\cdot s}{\longmapsto} Cln2\ |\ iCdh1 \\
    S_{13} : & Clb5\ |\ Cdh1 \stackrel{20\cdot s}{\longmapsto} iCdh1 \ |\ Clb5 \\
    S_{14} : & Clb5\ |\ iMcm1 \stackrel{0.25\cdot s}{\longmapsto} Mcm1 \ |\ Clb5 \\
    S_{15} : & Mcm1\, |\, (n)^L\rfloor (\tilde{y}.gB2.\tilde{x}\, |\, Y) \stackrel{1\cdot s}{\longmapsto} iMcm1\, |\,
(n)^L\rfloor (\tilde{y}.gB2.\tilde{x}\, |\, Y)\ |\ Clb2 \\
\end{array}
$

\normalsize

The accumulation of $Clb5$ is the event that triggers the transition from phase $S$ to phase $G_2$. We introduce the following vertical rule:
\small
\[T_2 : visualised_2 | Clb5^{mc(Clb5,3)} | stage_2 \stackrel{\infty}{\longmapsto} Clb5^{mc(Clb5,3)} | stage_3 | SPN \]
\normalsize
to instantaneously perform the transition from stage 2 to stage 3, after rule $R_2$ has visualised chromosome duplication.
Besides changing visual stage, the above rule also sends a signal (\emph{SPN}) to start metaphase spindle. This signal is needed to activate \emph{Cdc15} in mitosis. Since it is unknown how to relate the accumulation of this signal with biochemical reactions at molecular level, we assume that this signal is available since the beginning of phase $G_2$.

The cyclin-dependent kinase (CDK) \emph{Clb2-Cdc28} is the main controller of phase $G_2$ and $M$. This CDK activates \emph{Mcm1}, allowing \emph{Clb2} to accumulate. It also degrades \emph{MBF} and \emph{SBF}, stopping the transcription of \emph{Cln2} and \emph{Clb5}. Cyclin \emph{Cln2} is then degraded by \emph{SCF}, while the degradation of \emph{Clb5} is regulated by \emph{Cdc20} and \emph{APC} (\emph{Anaphase Promoting Complex}). \emph{Mcm1} stimulates gene \textit{gC20} to produce \emph{Cdc20} and \emph{APC} must be phosphorylated by \emph{Clb2-Cdc28} before it can bind with \emph{Cdc20}. During metaphase, the \emph{SPN} signal activates \emph{Cdc15}. Protein \emph{Cdc15} then phosphorylates \emph{Net1}, releasing \emph{Cdc14}. Protein \emph{Cdc14} is needed later in mitosis exit and also activates \emph{Cdh1}. Tumor suppressor \emph{Cdh1} is needed for \emph{Clb2} degradation.

\small
$
\begin{array}{ll}
    S_{16} : & Clb2\, |\, iMcm1 \stackrel{20\cdot s}{\longmapsto} Mcm1 \, |\, Clb2 \\
    S_{17} : & Clb2\, |\, MBF \stackrel{1\cdot s}{\longmapsto} Clb2\, |\, iMBF \\
    S_{18} : & Clb2\, |\, SBF \stackrel{1\cdot s}{\longmapsto} Clb2\, |\, iSBF \\
    S_{19} : & Mcm1\, |\, (n)^L\rfloor (\tilde{y}.gC20.\tilde{x}\, |\, Y) \stackrel{20\cdot s}{\longmapsto} iMcm1\, |\,
(n)^L\rfloor (\tilde{y}.gC20.\tilde{x}\, |\, Y)\ |\ Cdc20 \\
    S_{20} : & Clb2\, |\, APC \stackrel{20\cdot s}{\longmapsto} APC-P \, |\, Clb2 \\
    S_{21} : & APC-P\, |\, Cdc20 \stackrel{1\cdot s}{\longmapsto} APC-Cdc20 \\
    S_{22} : & SPN\, |\, iCdc15 \stackrel{0.25\cdot s}{\longmapsto} SPN \, |\, Cdc15 \\
    S_{23} : & Cdc15\, |\, Net1-Cdc14 \stackrel{0.25\cdot s}{\longmapsto} Net1 \, |\, Cdc14 \\
    S_{24} : & APC-Cdc20\, |\, Clb5 \stackrel{1\cdot s}{\longmapsto} APC \\
    S_{25} : & iCdh1 \, |\, Cdc14 \stackrel{0.25\cdot s}{\longmapsto} Cdh1\, |\, Cdc14 \\
\end{array}
$
\normalsize

The accumulation of $APC-Cdc20$ is the event that triggers the beginning of cytokinesis. We introduce the following vertical rule:
\small
\[T_3 : SPN | visualised_3 | APC-Cdc20^{mc(APC-Cdc20,4)} | stage_3 \stackrel{\infty}{\longmapsto} APC-Cdc20^{mc(APC-Cdc20,4)} | stage_4 \]
\normalsize
to instantaneously perform the transition from stage 3 to stage 4, after rule $R_3$ has visualised nucleus division. It also removes signal \emph{SPN} which is not needed anymore in cytokinesis, but whose degradation is not clearly understood in terms of biochemical reactions.

The main activity in the stage 4 is the degradation of \emph{Clb2} by \emph{APC} (with the help of \emph{Cdc20} or \emph{Cdh1}). Protein \emph{Cdc14} is also active in this stage, producing \emph{Sic1}, which is needed to return to phase $G_1$.

\small
$
\begin{array}{ll}
    S_{26} : & Cdc14 \stackrel{5\cdot s}{\longmapsto} Sic1\, |\, Cdc14 \\
    S_{27} : & APC-Cdc20\, |\, Clb2  \stackrel{1\cdot s}{\longmapsto} APC  \\
    S_{28} : & APC\, |\, Cdh1\, |\, Clb2  \stackrel{0.25\cdot s}{\longmapsto} APC\, |\, Cdh1
\end{array}
$
\normalsize

Finally, transition from cytokinesis to phase $G_1$ is triggered by the accumulation of $Sic1$. We introduce the following vertical rule:
\small
\[T_3 : visualised_4 | Sic1^{mc(Sic1,1)} | stage_4 \stackrel{\infty}{\longmapsto} Sic1^{mc(Sic1,1)} | stage_1 \]
\normalsize
to instantaneously perform the transition from stage 4 back to stage 1, after rule $R_4$ has visualised cell division.

\subsection{Algorithms for Simulation and Visualisation}\label{algo}

The model of cell cycle defined using the formal approach
introduced in Section~\ref{visual} and Section~\ref{molecular}
is used as the input for a variant of Gillespie's simulation algorithm
which extends the variant introduced by 
Basuki, Cerone and Milazzo~\cite{SCLSMaude}.
The extension consists of the selection of the cell
in which the current reaction has to occur
and the control of the visualisation.
\begin{description}
\item[Step 0] Input $M$ reactions $R_1,\ldots , R_M$, and $N$ values 
$X_1,\ldots , X_N$ representing the initial numbers of each
of $N$ kinds $B_1,\ldots , B_N$ of molecules.
Initialise time variable $t$ to 0.
Calculate propensities $a_j$, for $j=1,...,M$.
Calculate $\sum_{j=1}^{M} a_j$. 
\item[Step 1] If $visualised_l$ is present in the term
then visualise stage $l$ and execute vertical rule $T_l$.
\item[Step 2] If the space is fully occupied then stop simulation. 
Otherwise generate $r_1$ and calculate $\tau$. Increment $t$ by $\tau$.
\item[Step 3] Generate $r_2$ and calculate $(\mu,\sigma)$.
\item[Step 4] Execute $R_\mu$ inside cell $\sigma$.
Update $X_1,\ldots , X_N$ and $a_1,\ldots , a_M$ according to the
execution of $R_\mu$. 
\item[Step 5] Calculate $\sum_{j=1}^{M} a_j$.
Return to \textbf{Step 1}.                                                
\end{description}
 
Gillespie defines that the probability of reaction $R_j$ to occur is 
proportional to $a_j$, the \textit{propensity} of reaction $R_j$.
The propensity of reaction $R_j$ is the product of its rate $c_j$ and
the number $h_j$ of distinct combinations of reacting molecules.

Since reactions are confined within cells, we need to extend Gillespie's
algorithm to choose in which cell the chosen reaction $R_{\mu}$ should occur.
Let $C$ be the number of cells and $X_k^i$ be the number of molecules of kind
$B_k$ in the $i$-th cell.
We define $X_k = \sum_{i=1}^C X_k^i$.

Let $a_j^i$ be the propensity of reaction $R_j$ occurring inside the $i$-th cell.
Then $a_j^i$ is defined as the product of $c_j$ by the number $h_j^i$ of distinct
combinations of reacting molecules of $R_j$ within the $i$-th cell.
We define $a_j = \sum_{i=1}^C a_j^i$.
If $t$ is the current simulation time, then $t + \tau$ represents the time 
at which next reaction occurs, with $\tau$ exponentially distributed with 
parameter $a_0=\sum_{j=1}^{M}a_j$.
Time increment $\tau$,
the index $\mu$ of the reaction that occurs at time $t + \tau$ and
the index $\sigma$ of the cell in which such reaction occurs
are calculated as follows.
\begin{gather}
\tau = \frac{1}{a_0}ln(\frac{1}{r_1})
\label{taucalc}\\
(\mu,\sigma)=the\ integers\ f\! or\ which\ \sum_{j=1}^{\mu} \sum_{i=1}^{\sigma-1} a_j^i <\
r_2a_0\leq\ \sum_{j=1}^{\mu} \sum_{i=1}^{\sigma} a_j^i \label{muchoice}
\end{gather}
where $r_1$ and $r_2$ are two random real numbers which are uniformly distributed
over interval [0,1].

We assume the existence of a function, called $get\! pos()$, that
is responsible to find the correct position for a newborn cell and to resolve the spatial conflict that arises between cells. Naturally, the newborn cell should be attached to its parent cell. If we model the system in $n$-dimensional space, there are $2n$ positions for this newborn cell. Function $get\! pos()$ will find an empty position among these $2n$ positions. If it cannot find an empty position, it will choose one position and then push forward the other cells on that direction by one position. Since the space is limited by a sphere if the last cell is adjacent to the sphere boundary, it cannot be pushed forward. In this case $get\! pos()$ will search for any empty position for this cell. If no more empty position can be found, then the simulation must stop.

\begin{figure}[!htb]
    \begin{center}
        \includegraphics[width = 7 cm]{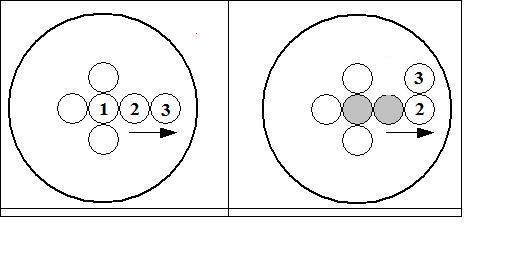}
        \caption{Application of $get\! pos()$}
       \label{fig:algogrid}
    \end{center}
\end{figure}

Figure \ref{fig:algogrid} shows an example in 2-dimensional space. In the left picture, cell number 1 is about to divide, but all neighboring positions are occupied by other cells. Then in the right picture $get\! pos()$ chooses the position of cell number 2 as its target and divides cell number 1 into two newborn cells (grey colour). It pushes cell number 2 toward cell number 3, but cannot push cell number 3 forward because of the boundary. Then cell number 3 is moved to an adjacent empty position, along a different direction.

After executing reaction $R_{\mu}$ (step 4 of the algorithm), we need to update the molecular populations and propensity functions that are affected by application of $R_{\mu}$. Gibson and Bruck \cite{gibbru00} define a data structure to support an extension of Gillespie's First Reaction Method. We use their notion of dependency graph in order to simplify the process of updating molecular populations and propensity functions. In this way we need to update them only if they are affected by the application of reaction $R_{\mu}$.

\subsection{Visualisation of Cell Cycle}
\begin{figure}[!htb]
    \begin{center}
        \includegraphics[width = 12 cm]{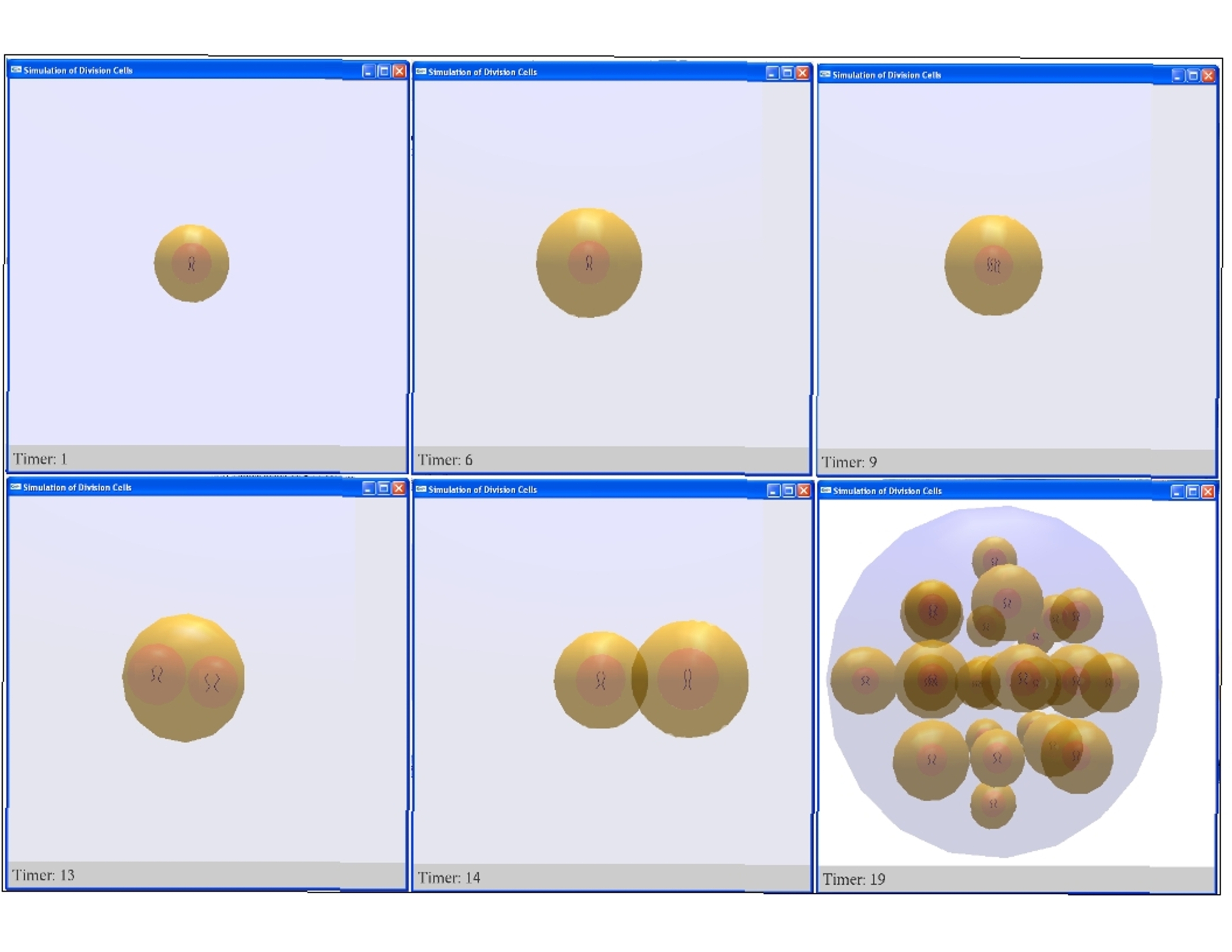}
        \caption{Visualisation of cell cycle by our tool}
       \label{fig:visual}
    \end{center}
\end{figure}

Figure~\ref{fig:visual} shows the visualisation of cell cycle in our tool. The picture on the top left shows the initial stage of the system, in which the sphere that limits the proliferation space only contains one small cell. This cell grows (top centre) and duplicates its chromosomes (top right). 
Nucleus division (bottom left) and cell division (bottom centre) complete the cell cycle. All newly born cells concurrently repeat the cell cycle and finally the simulation stops because the space is full of cells (bottom right).

The tool also allows to model the existence of a virus in the environment where cells proliferate. The virus can infect a cell through its membrane and duplicates itself inside the cell. We want to relate the virus infection on a cell with the cell's capability to proliferate. We model a virus that synthesises a protein able to degrade the growth factor receptor, which is needed in the growing phase of a cell. The virus can then move from the cell through the cell membrane back into the environment. From there it can then spread to another cell. We can classify the infection level of a cell based on the number of viruses inside it. We define a threshold $virusTH$ and use it to classify the level of infection of a cell into three classes as follows.
\begin{itemize}
    \item healthy cell, if no virus is inside the cell;
    \item lightly infected cell, if the number of viruses inside it is less than $virusTH$;
    \item severely infected cell, if the number of viruses inside it is greater than or equal to $virusTH$.
\end{itemize}

\begin{figure}[!htb]
    \begin{center}
        \includegraphics[width = 12 cm]{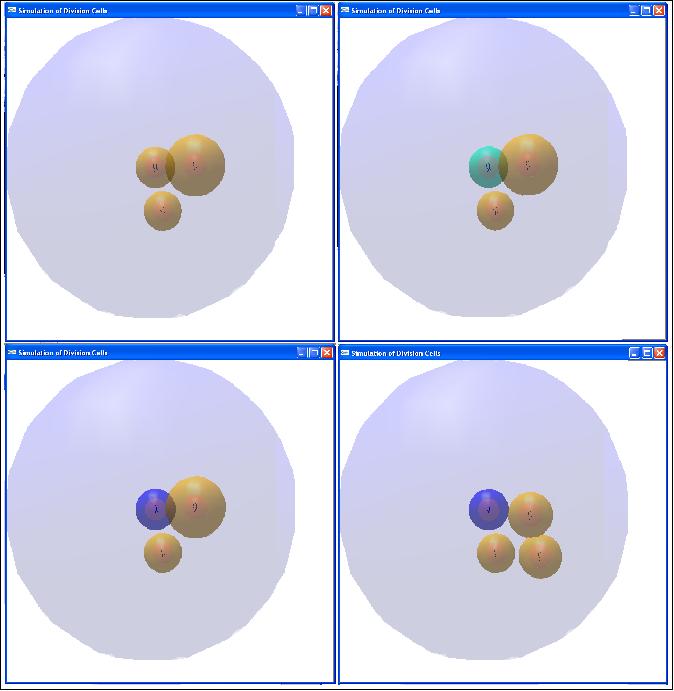}
        \caption{Visualisation of virus attack}
       \label{fig:virus}
    \end{center}
\end{figure}

To represent the infection level of a cell, we use colours in our visualisation. We colour healthy cells with orange, lightly infected cells with green and severely infected cells with blue. Figure~\ref{fig:virus} shows the visualisation of virus attack in our tool. The tool also supports the generation of a report file, describing the details of reactions occurring in the system. By combining our observation from the visualisation and the report, we can find interesting things about the model. For instance, in the visualisation we may observe that a severely infected cell can no longer proliferate, while other infected cells can still proliferate. We can later analyse the report generated by the tool and find out that the observed situation is related to the stage of the cell at the moment when it is infected by virus. If the cell is infected while it is still growing, this may cause the cell to loose its growing capability, which means that the cell can no longer proliferate. In contrast if the cell is infected after the growing phase is already over, it can still proliferate.

\section{Conclusions and Future Work}

We have defined an approach to model biological systems at different
levels of representation: a molecular level in which rewrite rules are
used to model biochemical reactions among molecules, and one or more visual
levels, in which rewrite rules define the dynamics of a higher level of
organisation of the biological system under analysis.
We have chosen Spatial CLS to model all representation levels in order to
formally describe, for every visual level, the spatial information needed
to quantitatively define all visual details that are believed to be essential
for an effective visualisation at that level.
Visual details whose quantitative definition is a purely aesthetic matter
are not included in the formal model and their quantitative definition is
left to the implementation.
Each molecular or visual level is defined by an initial term and a set of
rewrite rules, which we called horizontal rules since they only
refer to that level.
Each visual level is linked to the molecular level by a meta-level
of vertical rules, which also cater to lack of sufficient knowledge at
molecular level.

We have then presented the budding yeast cell cycle as the case study to
illustrate our approach, choosing the cellular level as the single visual
level.
Those visual details for which a quantitative characterisation is essential
for visualisation, such as the position and size of cells, have been formally
modelled, whereas irrelevant quantitative details, such as the position
within the nucleus of duplicated chromosomes, are not.

We have defined a variant of Gillespie's algorithm that exploits spatial
information to choose the cell in which next reaction has to occur according to the exponential distribution of reaction time,
and implemented the algorithm in a tool to illustrate the budding yeast cell
cycle case study.
In addition to the normal cell cycle, the tool supports the injection of a
virus in the environment where the cells proliferate.
The virus can enter a cell through its membrane, duplicate itself inside the
cell, and cause the degradation of the growth factor receptor.
The visualisation shows that some small infected cells can no longer proliferate
while big infected cells can still proliferate.
Although this specific situation is an obvious consequence of the degradation
effect the virus has on the growth factor receptor of the cell, it actually
illustrates that, in general, the visualisation of higher levels of organisation
has the potential to highlight important behaviours that would not
be captured through a formal analysis conducted at the molecular level.

As future work, we plan to extend the work by Basuki, Cerone and
Milazzo~\cite{SCLSMaude}, by implementing Spatial CLS within the MAUDE tool and parsing
the output generated by MAUDE to provide the appropriate input to our tool.
In this way, simulations performed using MAUDE could be visualised using the
tool.
Moreover, counterexamples generated by the model-checker associated with MAUDE
could be also visualised.
Finally we plan several extensions to the tool implementation, such as
\begin{itemize}
  \item the definition of several visual levels, corresponding to different
    magnifications of the biological system, with zooming capabilities to move
    from consecutive visual levels;
  \item the capabilities to save simulations and to re-visualise them at a later stage;
  \item the structuring and visualisation of the textual information, currently
    provided as a single report file, by attaching it to the visual objects for which
    it is relevant.
\end{itemize}
\bibliography{bio}
\bibliographystyle{plain}

\end{document}